# Accurate experimental ($p, \rho, T$) data of natural gas mixtures for the assessment of reference equations of state when dealing with hydrogen-enriched natural gas

Roberto Hernández-Gómez[1], Dirk Tuma[2], Daniel Lozano-Martín[1], César R. Chamorro[1]

[1] Grupo de Termodinámica y Calibración (TERMOCAL), Dpto. Ingeniería Energética y Fluidomecánica, Escuela de Ingenierías Industriales, Universidad de Valladolid, Paseo del Cauce, 59, E-47011 Valladolid, Spain.

[2] BAM Bundesanstalt für Materialforschung und -prüfung, D-12200 Berlin, Germany.

**Abstract**

The GERG-2008 equation of state is the approved ISO standard (ISO 20765-2) for the calculation of thermophysical properties of natural gas mixtures. The composition of natural gas can vary considerably due to the diversity of origin. Further diversification was generated by adding hydrogen, biogas, or other non-conventional energy gases. In this work, high-precision experimental ($p, \rho, T$) data for two gravimetrically prepared synthetic natural gas mixtures are reported. One mixture resembled a conventional natural gas of 11 components (11M) with a nominal mixture composition (amount-of-substance fraction) of 0.8845 for methane as the matrix compound with the other compounds being 0.005 for oxygen, 0.04 for nitrogen, 0.015 for carbon dioxide, 0.04 for ethane, 0.01 for propane, 0.002 each for n- and isobutane, and ultimately 0.0005 each for isopentane, n-pentane, and n-hexane. The other mixture was a 13-component hydrogen-enriched natural gas with a low calorific value featuring a nominal composition of 0.7885 for methane, 0.03 for hydrogen, 0.005 for helium, 0.12 for nitrogen, 0.04 for carbon dioxide, 0.0075 for ethane, 0.003 for propane, 0.002 each for n- and isobutane, and 0.0005 each for neopentane, isopentane, n-pentane, and n-hexane. Density measurements were performed in an isothermal operational mode at temperatures between 260 and 350 K and at pressures up to 20 MPa by using a single-sinker densimeter with magnetic suspension coupling. The data were compared with the corresponding densities calculated from both GERG-2008 and AGA8-DC92 equations of state to test their performance on real mixtures. The average absolute deviation from GERG-2008 (AGA8-DC92) is 0.027 % (0.078 %) for 11M and 0.095 % (0.062 %) for the 13-component $H_2$-enriched mixture, respectively. The corresponding maximum relative deviation from GERG-2008 (AGA8-DC92) amounts to 0.095 % (0.127 %) for 11M and 0.291 % (0.193 %) for the $H_2$-enriched mixture.

Keywords: hydrogen-enriched natural gas; gravimetric preparation; single-sinker densimeter; reference equations of state; GERG-2008; AGA8-DC92

\* Corresponding author e-mail: cescha@eii.uva.es. Tel.: +34 983423756.

ORCID:  Roberto Hernández-Gómez: 0000-0003-1830-1410

Dirk Tuma: 0000-0002-5713-9746

Daniel Lozano-Martín: 0000-0002-2286-0111

César R. Chamorro: 0000-0001-6902-3778

# 1. Introduction

The increasing global energy demand with its impact on mankind is a growing imperative for the development of alternative energy sources which truly fulfil the condition of sustainability. Initiated by the first oil crisis in 1973 [1], hydrogen is promoted as a sustainable alternative energy carrier in the implementation of energy policies that do not depend on fossil resources, or, in short, on carbon. The last years have seen a vibrant and continuously growing activity on all fields related. The concept of "decarbonisation" literally means to replace carbon atoms with hydrogen atoms. The book entitled "The Hydrogen Economy: Opportunities and Challenges" by Ball and Wietschel published in 2009 highlights hydrogen comprehensively from a technical, environmental, and socioeconomic perspective [2]. At present, the global economy still runs largely on fossil fuels. In a long-term perspective, fossil fuels must develop away from energy into the raw material for producing chemical products. Hydrogen, however, naturally occurs in a bonded form and thus has first to be released by using energy, ideally from renewable sources [3–7]. An illustrative example is the so-called "power-to-gas" technique, that is the use of surplus electrical power to produce a gas fuel, where hydrogen is made by electrolysis [8–10]. One hydrogen-related approach that has been put forward for making a transition towards a carbon-free economy is adding hydrogen into existing natural gas transport and distribution systems [11–15]. Capocelli and De Falco have put it in an introductory book chapter "Enriched methane is a ready solution for the transition towards the hydrogen economy" [16]; in other words, enriched methane is a pathway to introduce hydrogen in our established and consolidated energy infrastructure.

Apart from plain economic figures, various technical aspects over the entire chain, such as safety, availability, storage, metering, equipment and operational parameters, efficiency, as well as material characteristics, have to be successfully addressed towards a full affirmation. These research activities manifested in a notable number of studies. Some recently published papers which particularly address the aforementioned issues for enriched methane are: experiments of turbulent explosions, focusing on the influence on mixture reactivity by Li et al. [17] and explosion studies of methane–hydrogen mixtures by Lowesmith et al. [18] on safety; a study on transition pathways to future energy infrastructure by Maroufmashat and Fowler on availability [19]; development of measuring standards to support fiscal metering by van der Veen et al. on analysis and metering [20]; a comprehensive review on combustion engines for $H_2$-enriched natural gas by Mehra et al. [21], a study of low-calorific value coal gas combustion by Karyeyen and İlbaş [22,23], and the studies investigating the effects of composition on engine performance, combustion, and emission by Hora and Agarwal [24] and Cellek and Pınarbaşı [25], respectively, deal with equipment and operation.

Due to the low calorific value of hydrogen on a volumetric basis, a virtual hydrogen-spiked energy gas would require a gas with about 80 % of hydrogen to reduce the emission of carbon dioxide by 50 % [11]. The approved ISO standard (ISO 20765-2) for the calculation of thermophysical properties of established natural gas-type mixtures is the GERG-2008 equation of state [26,27]. The addition of hydrogen will contribute further diversification to the composition of those mixtures. The performance of equation-of-state (EoS) models, such as the GERG-2008 EoS or the AGA8-DC92 EoS developed by the American Gas Association [28], must be validated using consolidated volumetric ($pVT$) data of high accuracy. Real mixtures that are prepared by gravimetry according to ISO 6142-1 display the highest accuracy in composition and thus qualify best [29].

Prior to this work, Richter et al. have investigated three $H_2$-enriched natural gases [30]. The mixtures for this study were prepared from a 21-component high-calorific pipeline gas that was blended with hydrogen to a hydrogen content of approximately 5, 10, and 30 mol-%. Volumetric data were recorded at temperatures between 273 and 293 K and up to a maximum pressure of 8 MPa using a two-sinker densimeter. The experimental density data were compared with calculated values from both the GERG-2008 and the AGA8-DC92 EoS. For both EoS, the relative deviations in density remained below 0.05 % for the 5- and 10-% mixture but raised to a value of about 0.1 % for the 30-% mixture. The authors explained latter with larger errors in the composition analysis at a high hydrogen content.

In another volumetric study on mixtures, Atilhan, Aparicio, Hall, and co-authors have employed a ($H_2$-free) deepwater natural gas mixture with heavy hydrocarbon content (i.e., $C_{6+}$ up to n-$C_9$ > 0.2 mol-%) that were gravimetrically prepared and traceable to the National Institute of Standards and Technology NIST [31]. Since the authors intended to explore the limits of the equation-of-state models (both GERG-2008 and AGA8-DC92), the experiments were conducted at temperatures between 270 and 340 K and up to pressures of 35 MPa using a magnetic suspension densimeter. They found density deviations larger than 0.1 % especially for pressures < 10 MPa in the vicinity of the phase envelope for both models.

Chapoy and co-workers determined densities and speed-of-sound values of a synthetic natural gas (88 mol-% of methane) with certified composition at temperatures between 323 and 415 K and pressures up to 58 MPa [32] using a high-pressure and high-temperature vibrating-tube densimeter and a specially adapted in-house made acoustic cell. Additionally, an isochoric cell was employed to cover the entire operational range. In addition to GERG-2008 and AGA8-DC92 EoS, the (cubic) Peng-Robinson [33] and the Soave-Benedict-Webb-Rubin [34] EoS were used to evaluate the experimental data and to calculate various derivable thermodynamic properties. The Peng-Robinson EoS featured a maximum deviation in the predicted density of about 2.8 %, whereas the maximum deviation for the other models did not exceed 0.7 %.

This work provides new high-precision experimental ($p$, $\rho$, $T$) data for two gravimetrically prepared natural gas mixtures. The nominal composition of these mixtures is given in Table 1. Both mixtures also qualify as primary calibration standards. The first mixture resembled a conventional natural gas of 11 components and is labeled BAM-G 420 or 11M according to the specification given in the directive PTB-A 7.63 by the Physikalisch-Technische Bundesanstalt PTB [35]. The second mixture was a 13-component $H_2$-enriched natural gas mixture with a low calorific value to facilitate support to power-to-gas applications which was proposed by the Consultative Committee for Amount of Substance: Metrology in Chemistry and Biology (CCQM) of the Bureau International des Poids et Mesures (BIPM) for the interlaboratory key comparison K 118 [36]. Density measurements were performed at temperatures between 260 and 350 K and at pressures up to 20 MPa using a single-sinker densimeter with magnetic suspension coupling. 94 data points were recorded for 11M and 99 for the $H_2$-enriched gas mixture, respectively, in an isothermal operational mode at 260, 275, 300, 325, and 350 K. The data were compared with the corresponding densities calculated from GERG-2008 and AGA8-DC92 EoS, via REFPROP [37], and can thus give an assessment for the performance of these models on novel compositions of energy gases.

## 2. Experimental

### 2.1. Gas mixture preparation

Both natural gas mixtures were prepared gravimetrically by the Federal Institute for Materials Research and Testing (Bundesanstalt für Materialforschung und -prüfung, BAM) in Berlin, Germany, according to the procedures outlined in ISO 6142-1 [29].

[Table 1, Table 2]

The nominal composition of the two mixtures investigated as well as purity, supplier, molar mass, and critical parameters of the pure compounds are given in Table 1. Table 2 shows the cylinder identifiers, gravimetric composition, and the corresponding (absolute) expanded uncertainty ($k$ = 2) of the mixtures. The prepared mixtures were supplied in aluminum cylinders of a volume of 10 dm$^3$. All compounds were used without further purification, but information on impurities from the specification was considered in the mixture preparation.

Due to their particular composition, the preparation of the final mixtures required several premixtures and the transfer of liquid and liquefied compounds via high-pressure minicylinders. Liquid compounds of low volatility, such as n-hexane, were directly introduced by gas-tight glass syringes. A detailed chart for the individual preparation is given in the Supporting Information. In our method, pressure difference is the exclusive driving force for the mass transfer into the recipient cylinder. Optionally, heating had to be administered to create sufficient pressure differences as well as to avoid condensation and the assigned portions of liquid and liquefied compounds were always vaporized from the minicylinders during the transfer. The complete transfer of vapors and liquids from the tubing into the recipient cylinder required a purge gas that in turn was a pure gas or a mixture. The use of minicylinders for smaller mass portions ensured that the corresponding uncertainty of the weighed mass differences did not exceed the preset qualifying limits for the preparation of reference materials. During the filling procedure, the recipient cylinder freely stood on an electronic comparator balance (Sartorius LA 34000P-0CE, Sartorius AG, Göttingen, Germany, weighing range: 34 kg, readability: 0,1 g) to simultaneously monitor the gas feed stream. The exact mass of the gas added into the recipient cylinder was determined after each filling step using a high-precision mechanical balance (Voland model HCE 25, Voland Corp., New Rochelle NY, USA, weighing range: 25 kg, readability: 2.5 mg). The small masses, such as minicylinders and syringes, were determined on an electronic comparator balance (Sartorius CCE 2004, weighing range: 2500 g, readability: ≤ 0.2 mg). After the last filling step, each mixture was always homogenized by subsequent heating and rolling the cylinder.

Prior to density determination at the University of Valladolid, the two natural gas mixtures were validated at BAM by gas chromatography (GC) on a multichannel (12 in total for 6 trains) process analyzer (Siemens

MAXUM II, Siemens AG, Karlsruhe, Germany). The equipment was specifically designed and configured to analyze natural gas mixtures which can in general contain methane (as main compound), ethane, ethene, propane, propene, isobutane, n-butane, neopentane, isopentane, n-pentane, n-hexane, nitrogen, oxygen, carbon dioxide, carbon monoxide, helium, and hydrogen. The configuration as a process GC required the individual concentrations of the analyzable compounds to be within a predefined range. The GC operated in a single isothermal mode at 60 °C. Each channel was equipped with packed columns and a separate thermal conductivity detector (TCD). Helium and hydrogen were analyzed on a single channel operated by nitrogen as the carrier gas. The other channels were operated by helium and attributed to the analysis of a cluster of compounds. (There were also "non-analyzing" separate channels which served several backflush procedures and guard columns, respectively.) The compound clusters that shared one channel were combined as follows (with ascending retention times): methane, carbon dioxide, ethene, ethane — propane, propene — oxygen, nitrogen, carbon monoxide — isobutane, n-butane, neopentane, isopentane, n-pentane — n-hexane.

[Table 3]

The validation was performed according to ISO 12963, preferably by a two-point bracketing method using independently prepared calibration gases [38]. The 11M mixture was frequently used as calibration gas (BAM-G 420) in routine certifications, the $H_2$-enriched natural gas mixture was validated during the validation campaign for the samples to be distributed to the participants in the interlaboratory key comparison K 118 in which BAM acted as pilot lab. The results of the GC analysis and the corresponding (gravimetric) composition of the appropriate gas mixtures used for validation go with Table 3. There, a mixture regularly employed in certification analyses by a direct match method is given for the 11M mixture and both mixtures to perform a two-point bracketing method for the $H_2$-enriched natural gas mixture, respectively. The deviations between gravimetric composition and that from GC analysis were sufficiently low to pass the criteria for certification.

## 2.2. Apparatus and method

The volumetric ($p$, $\rho$, $T$) data were recorded with a single-sinker densimeter with magnetic suspension coupling. That particular method operates on the Archimedes' principle. A magnetic suspension coupling system enables to determine the buoyancy force on a sinker immersed in the medium so that density values over a large range of temperature and pressure can be measured with high accuracy [39]. The design had been adapted and optimized for density measurements of both pure gases and gaseous mixtures by researchers from the University of Bochum, Germany [40,41]. For the specific details of the apparatus used in this work the reader is referred to the paper by Mondéjar et al. and the references cited therein [42].

The sinker used here had a cylindrical shape and was made of silicon with a mass of 61.59181 ± 0.00016 g and a corresponding volume of 26.444 ± 0.003 cm$^3$ ($k = 2$), determined at $T = 293.05$ K and $p = 1.01134$ bar. The device operated with a load compensation system that consisted of two calibrated masses (provided by Rubotherm GmbH, Bochum, Germany) made of tantalum and titanium, respectively, that have approximately the same volume (4.9 cm$^3$) but different masses. Mass and volume of these mass pieces underwent a calibration at the Mass Division of the Spanish National Metrology Institute (Centro Español de Metrología, CEM) prior to their use [43]. The weight difference of both masses resembles that of the sinker. The characteristic load compensation allows for running the measurements near the zero point of the balance where the effect of air buoyancy becomes negligible. A detailed description of operational procedures and data reduction to obtain the final result for the density value is given in the papers by Mondéjar et al. [42] for our equipment and by Richter et al. [44] and McLinden [45] on general aspects. Principally, density $\rho$ is calculated using equation 1

$$\rho = \frac{(m_{S0} - m_{Sf})}{V_S(T,p)} \tag{1}$$

where the numerator represents the buoyancy force that is exerted on the sinker. The mass $m_{S0}$ stands for the weighing result of the sinker in vacuum, $m_{Sf}$ for the corresponding result in a pressurized medium. The weighing operation is done with a high-precision electronic microbalance (Mettler Toledo XPE205DR, Mettler Toledo GmbH, Gießen, Germany, normal weighing range: 81 g, readability: 0.01 mg, extended weighing range: 220 g, readability: 0.1 mg). The denominator $V_S(T, p)$ of equation 1 is the pressure- and temperature-dependent volume of the sinker immersed in the respective medium. The temperature in the cell is determined by two platinum resistance thermometers (S1059PJ5X6, Minco Products, Inc.,

Minneapolis MN, USA) and a reference resistance which are connected with an AC comparator resistance bridge (F700, Automatic Systems Laboratories, Redhill, England). The pressure is recorded by two pressure transducers for the entire operational range, namely a Paroscientific 2500A-101 for pressures from 0 to 3 MPa and a Paroscientific 43KR-HHT-101 (Paroscientific Inc., Redmond WA, USA) for elevated pressures up to 20 MPa. To achieve a higher accuracy, evaluation of data from the single-sinker technique requires a correction for two effects that occur, i.e., the force transmission error (FTE) due to magnetic coupling [46] and absorption of gas molecules on the surface inside the cell and sinker [47]. The FTE is a combination of two effects, namely an apparatus effect and a medium-specific effect. The apparatus effect can be eliminated by simply determining the sinker mass in vacuum $m_{S0}$ after finishing an isotherm (see also section 2.3 Experimental). However, the medium-specific effect results from the magnetic susceptibilities ($\chi$) of the mixtures. Literature recommends a consideration only for strongly paramagnetic fluids [46], such as oxygen or oxygen mixtures. The susceptibility value of the studied gas mixtures, composed mainly of methane, estimated with the additive law proposed by Bitter [48], is three orders of magnitude lower than the magnetic susceptibility of oxygen ($\chi_{O2} = 1.78 \cdot 10^{-6}$). Additionally, the sinker density also contributes to the medium-specific effect, with low-density sinkers being preferable [46]. In our work, the silicon sinker is of low density ($\rho_{Si} = 2329$ kg·m$^{-3}$) compared to the titanium sinkers ($\rho_{Ti} = 4506$ kg·m$^{-3}$) used by other single sinker densimeters. The low susceptibility values of the studied gas mixtures in conjunction with the low density of the silicon sinker make a medium-specific correction negligible that was therefore discarded. The influence of sorption effects inside the measuring cell is less defined and thus more difficult to specify due to particular interactions between the medium and the surface. Other researchers working with similar techniques reported errors in density up to 0.1 % from such phenomena [47]. The best and approved way to minimize that effect is a procedure of several alternating flushing and evacuating the measuring cell before the actual measurement is started. The residence time inside the cell of the mixture to be investigated did never exceed a period of 40 hours. Another viable way to obtain more information on those sorption effects was a specific sorption test for the particular gas mixture in the same way as executed in previous studies [49,50].

**2.3. Experimental**

The volumetric ($p$, $\rho$, $T$) data were recorded as isotherms. Both mixtures were investigated at temperatures of 260, 275, 300, 325, and 350 K and up to a maximum pressure of 20 MPa. Figure 1 illustrates the recorded data as coordinates in a $p$, $T$-diagram together with the saturation curve calculated using the GERG-2008 EoS [26]. The $p$, $T$-range where the GERG-2008 EoS can be applied and the region of technical relevance are also shown in Figure 1. To ensure an error-free operation of the device, nitrogen as a reference fluid was investigated at selected conditions over the entire operational range [50]. During a measurement, the pressure was reduced in intervals of 1 MPa starting from 20 MPa down to 1 MPa for each isotherm. Thirty repeated measurements were processed into one definite data point. The closing procedure of each isothermal measuring series was the determination of the true sinker mass under vacuum conditions $m_{S0}$ to detect any misalignment from the magnetic suspension couplings.

[Figure 1]

**2.4. Uncertainty analysis**

The evaluation of uncertainties (all given as expanded uncertainties, i.e., $k = 2$) followed the same procedure as in previous studies [42]. The uncertainty in temperature amounted to less than 4 mK, the uncertainty in pressure depended on the range covered by the individual transducer. Equation 2 gives the uncertainty relation for the high-pressure (3 to 20 MPa) transducer, and equation 3 for the low-pressure transducer (0 to 3 MPa), respectively. The uncertainty in pressure in both mixtures remained below 0.005 MPa.

$$U(p) / \text{MPa} = 75 \cdot 10^{-6} \cdot p / \text{MPa} + 3.5 \cdot 10^{-3} \qquad (2)$$

$$U(p) / \text{MPa} = 60 \cdot 10^{-6} \cdot p / \text{MPa} + 1.7 \cdot 10^{-3} \qquad (3)$$

The uncertainty of the density value was calculated by executing the uncertainty propagation law on equation 1 according to the procedures given in the Guide to the Expression of Uncertainty in Measurement (GUM) [51]. From equation 1, the true sinker mass $m_{S0}$, the apparent sinker mass in the medium $m_{Sf}$, and the sinker volume $V_S$ ($T$, $p$) contribute to the uncertainty. Additionally, the uncertainty of the apparent sinker mass $m_{Sf}$ includes the entry from calibration, resolution, repeatability, and balance drift. Since the sinker volume $V_S$ is affected by temperature and pressure, equation 4 describes the effect of thermal expansion and pressure distortion as a function of density.

$$U(\rho) / \text{kg} \cdot \text{m}^{-3} = 1.1 \cdot 10^{-4} \cdot \rho / \text{kg} \cdot \text{m}^{-3} + 2.3 \cdot 10^{-2} \quad (4)$$

The overall expanded uncertainty in density $U_T(\rho)$ includes the uncertainties of density, temperature, pressure, and ultimately the composition of the gas mixture, see equation 5.

$$U_T(\rho) = 2 \cdot \left[ u(\rho)^2 + \left(\left(\frac{\partial \rho}{\partial p}\right)_{T,x} \cdot u(p)\right)^2 + \left(\left(\frac{\partial \rho}{\partial T}\right)_{p,x} \cdot u(T)\right)^2 + \sum_i \left(\left(\frac{\partial \rho}{\partial x_i}\right)_{T,p,x_j \neq x_i} \cdot u(x_i)\right)^2 \right]^{0.5} \quad (5)$$

In equation 5, apart from temperature $T$ and pressure $p$, $x_i$ is the amount-of-substance (mole) fraction of each individual mixture component. Partial derivatives of equation 5 may be calculated from GERG-2008 EoSe via REFPROP [37]. The individual uncertainty contributions are summarized in Table 4.

[Table 4]

## 3. Results and discussion

The experimental volumetric ($p$, $\rho$, $T$) data measured for the two natural gas mixtures during this work are given in Table 5 for the 11M (BAM-G 420) mixture and Table 6 for the $H_2$-enriched mixture together with the corresponding expanded uncertainty in density from equation 4 and expressed both in density units (kg m$^{-3}$, i.e., absolute value) and as percentage of the measured density (i.e., relative value). The experimental data were compared to the corresponding density data calculated from the GERG-2008 and AGA8-DC92 EoS. Two columns in the Tables 5 and 6 represent the relative deviations between experimental and calculated data and the corresponding data are plotted in Figures 2 to 5.

[Tables 5, 6      Figures 2, 3, 4, 5]

Figures 2 and 3 show the relative deviations of experimental density data ($\rho_{\text{exp}}$) from density data calculated by the GERG-2008 ($\rho_{\text{GERG}}$) and AGA8-DC92 ($\rho_{\text{AGA}}$) EoS versus pressure $p$ at a constant temperature $T$ for the 11M natural gas mixture. Figures 4 and 5 represent the equivalent for the $H_2$-enriched natural gas mixture.

The intrinsic uncertainty of the GERG-2008 EoS in the gas-phase region over the temperature range from 250 to 450 K and at pressures up to 35 MPa amounts to 0.1 % in density [26].

A look at the Figures 2 to 5 shows that a negative deviation of the calculated density compared to the experimental density dominates for both gas mixtures and both equation-of-state models. The relative deviations approximate to zero towards low pressures as it is expected from the ideal gas limit and is a prove of the stated composition of the mixtures and of its stability. For the 11M mixture processed by the GERG-2008 EoS shown in Figure 2, the data follow a relatively flat sinusoid curve but all data remain within the 0.1-% margin, with the largest deviation for the lowest temperature $T = 260$ K. The effect becomes stronger and more temperature dependent at pressures $p > 10$ MPa. The results for AGA8-DC92 on the 11M mixture in Figure 3 look differently. The deviations are all negative and the deviation increases slowly but monotonously towards increasing pressure. A comparison of the performance of both equation-of-state models shows that GERG-2008 performs better on 11M than AGA8-DC92, but the gain diminishes when going towards lower temperatures. The results for the $H_2$-enriched gas mixtures are apparently different. The GERG-2008 EoS shown in Figure 4 resulted in a quite similar pattern as for the 11M mixture, but more strongly pronounced. The deviation becomes steeper at about 5 MPa and goes through a maximum of about 0.30 % at around 16 MPa. There is a visible temperature dependence. In contrast, the AGA8-DC92, shown in Figure 5, shows a smaller deviation and a less pronounced temperature dependence until a pressure of 15 MPa where a steeper deviation begins. As expected, the performance on the $H_2$-enriched mixture is generally poorer and, surprisingly, the AGA8-DC92 gives better results.

[Table 7]

The findings illustrated in Figures 2 to 5 were further evaluated by statistical parameters that were already applied in previous studies [50] and are given in Table 7. Equation 6 defines the average absolute deviation *AAD*, equation 7 the so-called *Bias* that quantifies the average deviation of the data set, and equation 8 represents the root mean square *RMS*. The subscript "EoS" is replaced by "GERG" or "AGA" to denote the applied equation-of-state model in the corresponding Tables and Figures.

$$AAD = \frac{1}{n}\sum_{i=1}^{n}\left|10^2 \cdot \frac{\rho_{i,\exp} - \rho_{i,EoS}}{\rho_{i,EoS}}\right| \qquad (6)$$

$$Bias = \frac{1}{n}\sum_{i=1}^{n}\left(10^2 \cdot \frac{\rho_{i,\exp} - \rho_{i,EoS}}{\rho_{i,EoS}}\right) \qquad (7)$$

$$RMS = \sqrt{\frac{1}{n}\sum_{i=1}^{n}\left(10^2 \cdot \frac{\rho_{i,\exp} - \rho_{i,EoS}}{\rho_{i,EoS}}\right)^2} \qquad (8)$$

Additionally, *MaxD* stands for the maximum relative deviation in the considered data set given as absolute value. For the 11M natural gas the *AAD* of 0.027 from the GERG-2008 EoS was lower than the corresponding value of 0.073 from the AGA8-DC92 EoS. A similar relation was found for the *RMS* values, and – expectedly – GERG-2008 also produced a lower value for *MaxD*. However, the result for the H$_2$-enriched natural gas was the opposite. Here, the application of AGA8-DC92 gave a lower *AAD* value of 0.062 compared to 0.095 for GERG-2008. Consequently, AGA8-DC92 produced lower values also for *RMS* and *MaxD*, respectively.

The statistical analysis according to the equations 6 to 8 was applied on some selected recently published literature data that had executed both equation-of-state models on their results for comparison [30–32]. The mixtures studied by Richter and co-workers were H$_2$-enriched, 21-component high-calorific gases with 5 (NG1), 10 (NG2), and 30 mol-% hydrogen (NG3) originating from a pipeline. The characteristic deviation pattern found in their study looks differently to our results. Data processing with GERG-2008 resulted mostly in a positive deviation that goes through a flat maximum between 4 and 6 MPa for all three mixtures. The AGA8-DC92 EoS produced more negative deviations for the respective coordinates, but the values for *AAD*, *Bias*, and *RMS* turned out to be smaller than the corresponding values for the GERG-2008 EoS except for NG3, i.e., the mixture with the highest hydrogen content of 30 mol-%.

The mixtures studies by Atilhan et al. contained heavier alkanes but no hydrogen [31]. The methane content of M88C1 (88 mol-%) was smaller than for M94C1 (94 mol-%) with that difference being merely compensated by higher amounts of ethane (5.8 mol-% for M88C1 and 1.9 mol-% for M94C1) and propane (3.3 mol-% for M88C1 and 1.8 mol-% for M94C1), respectively. In both mixtures, carbon dioxide (1.5 mol-%) and nitrogen (2.5 mol-%) were at approximately similar concentrations. Data analysis with the GERG-2008 EoS resulted in a sinusoid function of both positive (at $p > 25$ MPa) and negative (at $p < 25$ MPa) deviations, with no significant temperature dependence for M88C1. The corresponding diagram of M94C1, however, shows mostly negative deviations that turn into positive values not before $p > 30$ MPa. A temperature dependence clearly manifested in the largest deviation at low temperatures. Data processing of M88C1 with the AGA8-DC92 EoS gave a similar pattern but with a stronger temperature dependence than for GERG-2008; the results for M94C1 display a good coincidence with minor deviations only. An assessment of the statistical parameters given in Table 7 states a better performance of GERG-2008 on M88C1 and of AGA8-DC92 on M94C1. However, the difference between the two models remains rather small.

Ultimately, Ahmadi et al. investigated a rather "ordinary" natural gas mixture of 7 compounds (methane: 88 mol-%, ethane: 6.0 mol-%, propane: 2.0 mol-%, n-butane: 0.3 mol-%, isobutane: 0.2 mol-%, carbon dioxide: 2.0 mol-%, nitrogen: 1.5 mol-%) [32]. A characteristic feature of their results (the detailed plot was provided for the GERG-2008 EoS only) is that scattering and (significant) deviation starts at about pressures of 15 MPa and further increases towards lower pressure. When this area is not considered in the analysis, experimental and calculated data show good coincidence even at the highest pressures investigated.

## 4. Conclusions

In this work, new high-precision experimental ($p$, $\rho$, $T$) data for two multicomponent natural gas mixtures were recorded. The gas mixtures for this study mimic real natural gas mixtures, one being a conventional 11-component gas, the other a 13-component $H_2$-enriched mixture proposed by CCQM. They were prepared by gravimetry to create a metrologically traceable mixture which qualifies as reference material. The experiments were done using a single-sinker densimeter with magnetic suspension coupling in a temperature range between 260 and 350 K and up to pressures of 20 MPa. Subsequently, the new data were compared with the corresponding densities calculated with the two established reference EoS for natural gases, namely GERG-2008 and AGA8-DC92. The actual performance of the equation-of-state model can be tested using consolidated data of real mixtures. The application of the two models on the density data from the present study resulted in an average absolute deviation that remained below 0.1 % for both natural gas mixtures and both models, respectively, for the $p$, $T$-range investigated.

For the 11M natural gas mixture processed by the GERG-2008 EoS all data remain within the 0.1-% margin, with the largest deviation for the lowest temperature $T$ = 260 K. The results for AGA8-DC92 on the 11M mixture remain also within the 0.1-% margin, except at higher pressures, where the differences with the EoS are slightly bigger than 0.1 %. A comparison of the performance of both equation-of-state models shows that GERG-2008 performs better on 11M than AGA8-DC92.

The results for the $H_2$-enriched gas mixtures are different. The deviations of experimental data from the GERG-2008 EoS are not within the 0.1-% margin at high pressures, this pressure limit being lower for lower temperatures (17 MPa for 350 K, and only 9 MPa for 260 K). The deviation may be as high as 0.30 % at 260 K and 16 MPa. In contrast, the deviations from the AGA8-DC92 EoS are smaller, and are found outside the 0.1 %-margin only for the two lower temperatures measured (i.e., 260 K and 275 K) and pressures above 16 MPa. The maximum deviation is near 0.2 % at 260 K and 20 MPa. In this case, the performance of AGA8-DC92 is better than that of GERG 2008.

Generally speaking, the performance of the AGA8-DC92 EoS is satisfactory when dealing both with natural gas mixtures and with hydrogen-enriched natural gas mixtures, giving in both cases similar results. Otherwise, the GERG-2008 EoS displays a better performance than AGA8-DC92 when applied on with natural gas-type mixtures, but presents higher deviations for hydrogen-enriched natural gas mixtures at low temperatures and high pressures.

Nevertheless, our results only cover a limited $p$, $T$-sector compared to the approved overall operational range of both equation-of-state models. As the literature comparison shows, a different behavior can occur at other temperatures and pressures, which neither can directly be drawn from an extrapolation nor is easily predictable. Improvement of the theoretical models designed for natural gas, especially for their use with hydrogen-enriched natural gas mixtures, will need – consolidate – experimental thermodynamic properties of binary mixtures of hydrogen with the main components of natural gas [50, 52]. The user community needs to know about the generally accepted limits of the models and their performance in the vicinity of these limits for different mixtures. Our study was done at conditions that are often encountered during applications, and the density data obtained from samples of metrological traceability were processed with both reference EoS, which produced results of good quality.

**Acknowledgments**

Support for this work came from the projects 'Caracterizacion de gases energeticos sostenibles (biogas e hidrogeno), producidos con recursos renovables biomásicos y eólicos, para su incorporación a la red de gas natural', ENE2017-88474-R, funded by the Spanish Government and 'Revalorización de recursos renovables regionales biomásicos y eólicos para la producción de gases energéticos sostenibles (biogás e hidrógeno) y su incorporación en la red de gas natural', VA035U16 and 'UIC-114' from the Junta de Castilla y León.

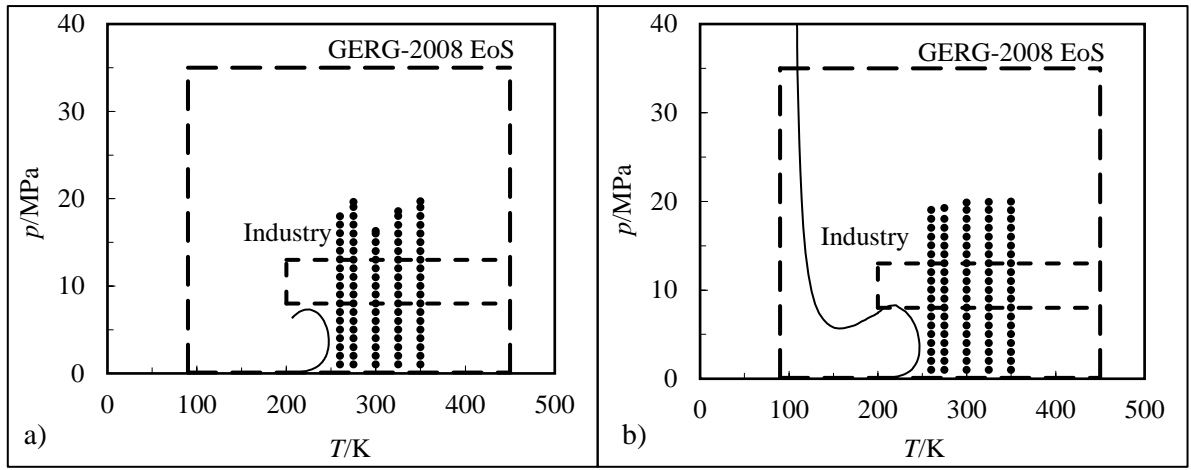

**Figure 1.** *p, T*-phase diagram showing the experimental points measured (●) and the calculated saturation curve for the a) 11M natural gas-like mixture and b) $H_2$-enriched natural gas mixture. The marked temperature and pressure ranges represent the validity of the GERG-2008 EoS (dashed line) and the area of interest for the transport and compression of gas fuels (thin dashed line).

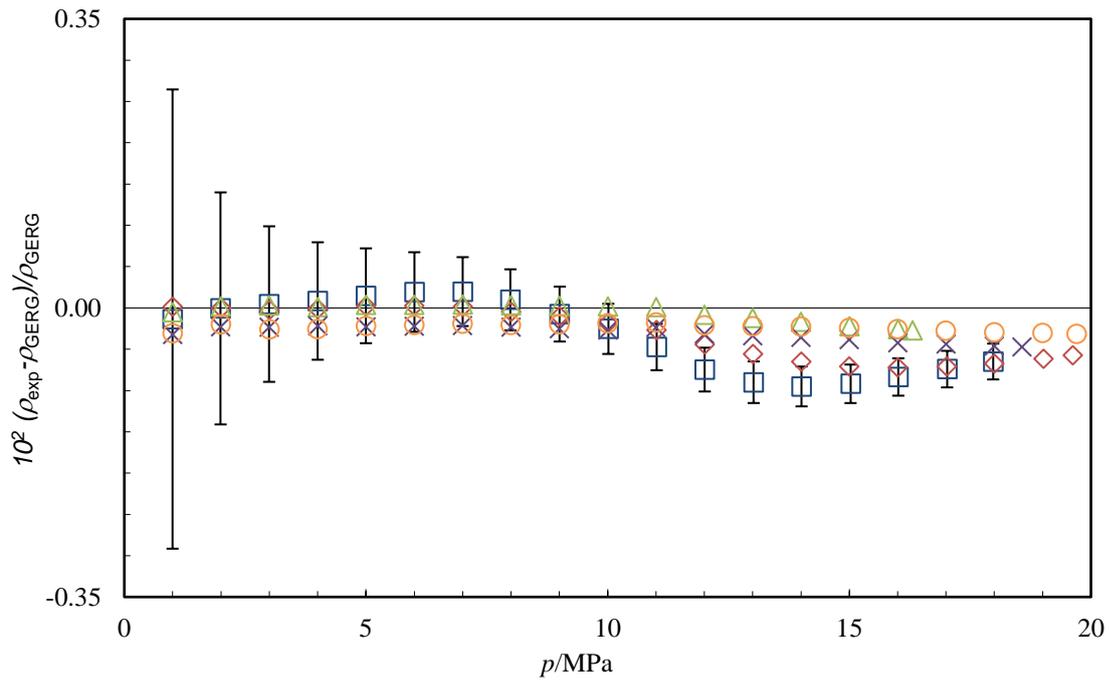

**Figure 2.** Relative deviations in density of experimental ($p$, $\rho$, $T$) data of the 11M natural gas-like mixture $\rho_{exp}$ from density values calculated from the GERG-2008 EoS $\rho_{GERG}$ versus pressure $p$: □, $T$ = 260 K; ◇, $T$ = 275 K; △, $T$ = 300 K; ×, $T$ = 325 K; ○, $T$ = 350 K. Error bars on the 260-K isotherm indicate the expanded uncertainty ($k$ = 2) of the experimental density data calculated with equation 4.

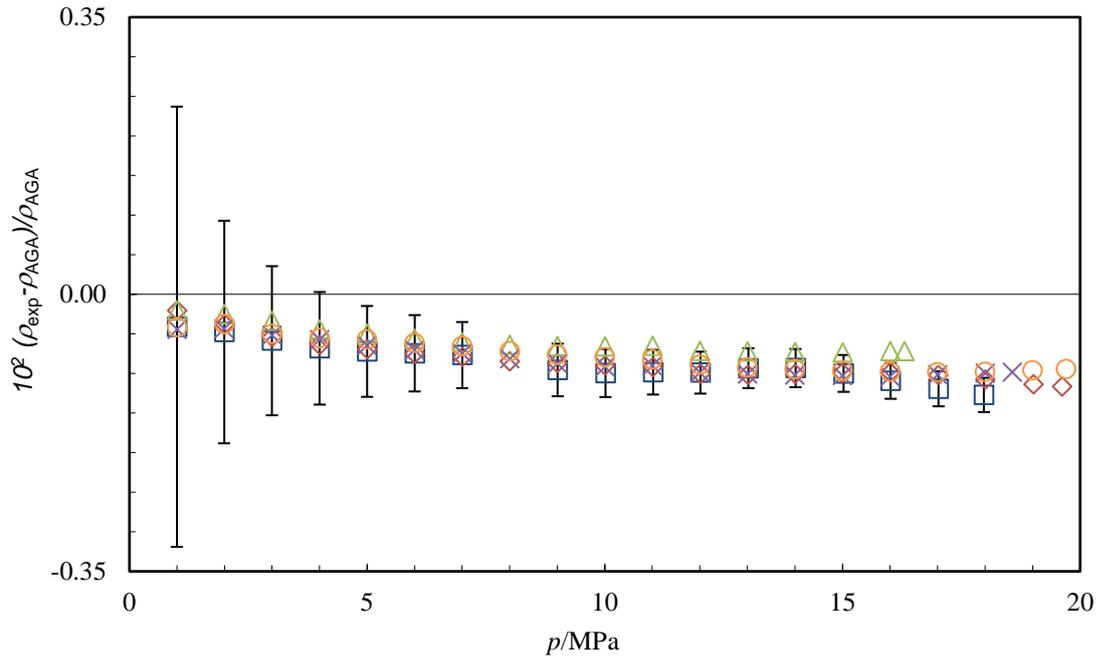

**Figure 3.** Relative deviations in density of experimental ($p$, $\rho$, $T$) data of the 11M natural gas-like mixture $\rho_{exp}$ from density values calculated from the AGA8-DC92 EoS $\rho_{AGA}$ versus pressure $p$: □, $T = 260$ K; ◇, $T = 275$ K; △, $T = 300$ K; ×, $T = 325$ K; ○, $T = 350$ K. Error bars on the 260-K isotherm indicate the expanded uncertainty ($k = 2$) of the experimental density data calculated with equation 4.

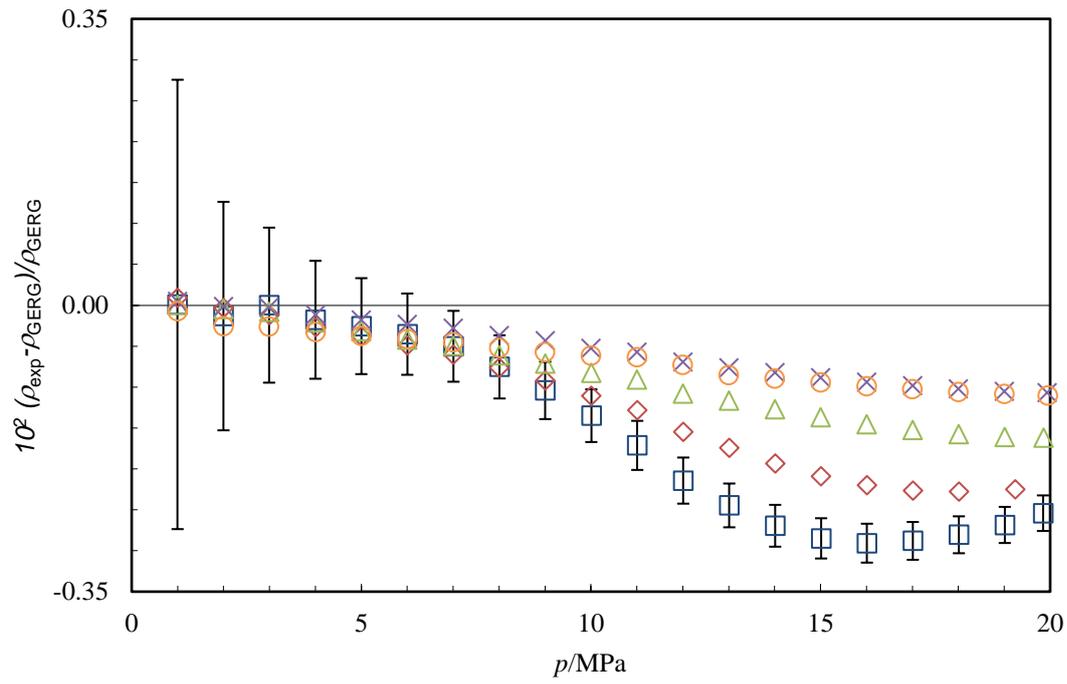

**Figure 4.** Relative deviations in density of experimental ($p$, $\rho$, $T$) data of the H$_2$-enriched natural gas mixture $\rho_{\text{exp}}$ from density values calculated from the GERG-2008 EoS $\rho_{\text{GERG}}$ versus pressure $p$: □, $T$ = 260 K; ◇, $T$ = 275 K; △, $T$ = 300 K; ×, $T$ = 325 K; ○, $T$ = 350 K. Error bars on the 260-K isotherm indicate the expanded uncertainty ($k$ = 2) of the experimental density data calculated with equation 4.

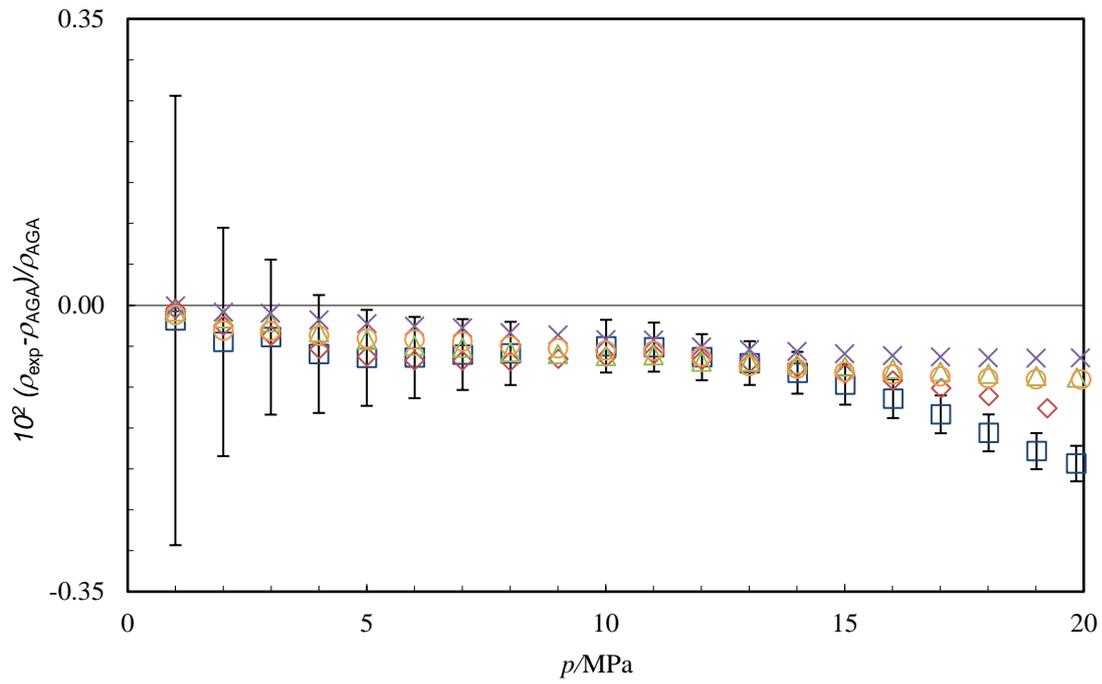

**Figure 5.** Relative deviations in density of experimental ($p$, $\rho$, $T$) data of the H$_2$-enriched natural gas mixture $\rho_{\text{exp}}$ from density values calculated from the AGA8-DC92 EoS $\rho_{\text{AGA}}$ versus pressure $p$: □, $T$ = 260 K; ◇, $T$ = 275 K; △, $T$ = 300 K; ×, $T$ = 325 K; ○, $T$ = 350 K. Error bars on the 260-K isotherm indicate the expanded uncertainty ($k$ = 2) of the experimental density data calculated with equation 4.

**Table 1.** Nominal composition of the two gas mixtures, purity, supplier, molar mass, and critical parameters of the individual components.

| Components | Synthetic natural gas 11M (BAM-G 420) | | H$_2$-enriched natural gas CCQM K 118 | | Purity (mol-%) | Supplier | $M$ / g·mol$^{-1}$ | Critical parameters [a] | |
|---|---|---|---|---|---|---|---|---|---|
| | $x_i$ (10$^{-2}$ mol·mol$^{-1}$) | $U(x_i)$ ($k=2$) [b] (10$^{-2}$ mol·mol$^{-1}$) | $x_i$ (10$^{-2}$ mol·mol$^{-1}$) | $U(x_i)$ ($k=2$) [c] (10$^{-2}$ mol·mol$^{-1}$) | | | | $T_c$ / K | $p_c$ / MPa |
| Methane | 88.45 | 0.044 | 78.85 | – | ≥ 99.9995 | Linde [d] | 16.043 | 190.56 | 4.60 |
| Ethane | 4.0 | 0.016 | 0.75 | – | ≥ 99.999 | Matheson [e] | 30.069 | 305.32 | 4.87 |
| Propane | 1.0 | 0.004 | 0.30 | – | ≥ 99.999 | Scott [f] | 44.096 | 369.89 | 4.25 |
| Butane | 0.2 | 0.001 | 0.20 | – | ≥ 99.95 | Scott [g] | 58.122 | 425.13 | 3.80 |
| Isobutane | 0.2 | 0.001 | 0.20 | – | ≥ 99.98 | Scott [h] | 58.122 | 407.81 | 3.63 |
| Pentane | 0.05 | 0.00025 | 0.05 | – | ≥ 99.7 | Sigma [i] | 72.149 | 469.70 | 3.37 |
| Isopentane | 0.05 | 0.00025 | 0.05 | – | ≥ 99.7 | Sigma [i] | 72.149 | 460.35 | 3.38 |
| Neopentane | – | – | 0.05 | – | ≥ 99.0 | Linde [d] | 72.149 | 433.74 | 3.20 |
| Hexane | 0.05 | 0.00025 | 0.05 | – | ≥ 99.7 | Sigma [i] | 86.175 | 507.82 | 3.03 |
| Carbon dioxide | 1.5 | 0.008 | 4.0 | – | ≥ 99.9995 | Air Liquide [j] | 44.010 | 304.13 | 7.38 |
| Nitrogen | 4.0 | 0.012 | 12.0 | – | ≥ 99.9995 | Linde [d] | 28.013 | 126.19 | 3.39 |
| Oxygen | 0.5 | 0.003 | – | – | ≥ 99.9999 | Westfalen [k] | 31.999 | 154.58 | 5.04 |
| Helium | – | – | 0.50 | – | ≥ 99.9995 | Linde [d] | 4.003 | 126.19 | 3.40 |
| Hydrogen | – | – | 3.0 | – | ≥ 99.9999 | Linde [d] | 2.016 | 33.15 | 1.30 |

[a] Critical parameters were obtained by using the default equation for each substance in REFPROP software [37].
[b] The given uncertainty is certified by BAM on request according to the PTB document PTB-A 7.63 [35].
[c] CCQM agreed on a nominal target composition only [36].
[d] Linde AG, Unterschleißheim, Germany.
[e] Matheson Tri-Gas, Inc., Montgomeryville PA, USA.
[f] Scott Specialty Gases BV, Breda, The Netherlands.
[g] Scott UK, Newcastle-under-Lyme, UK.
[h] Scott Specialty Gases, Inc., Plumsteadville PA, USA.
[i] Sigma-Aldrich Chemie GmbH, Steinheim, Germany.
[j] Air Liquide AG, Düsseldorf, Germany.
[k] Westfalen AG, Münster, Germany.

**Table 2**. Gravimetric composition of the natural gas mixtures studied including information on impurities from the supplier. Impurity compounds are marked in *italic* type.

| Component | 11M (BAM-G 420) natural gas BAM no.: C49358-090825 | | H$_2$-enriched natural gas CCQM K 118 BAM no.: 8099-160905 | |
| --- | --- | --- | --- | --- |
| | $x_i$ / ($10^{-2}$ mol·mol$^{-1}$) | $U(x_i)$ ($k=2$) / ($10^{-2}$ mol·mol$^{-1}$) | $x_i$ / ($10^{-2}$ mol·mol$^{-1}$) | $U(x_i)$ ($k=2$) / ($10^{-2}$ mol·mol$^{-1}$) |
| Methane | 87.663627 | 0.003485 | 78.821237 | 0.003809 |
| Ethane | 4.225210 | 0.000462 | 0.757359 | 0.000158 |
| Propane | 1.049010 | 0.002110 | 0.297078 | 0.000089 |
| Butane | 0.212654 | 0.000101 | 0.200439 | 0.000098 |
| Isobutane | 0.210325 | 0.000084 | 0.197953 | 0.000035 |
| Pentane | 0.051829 | 0.000027 | 0.050134 | 0.000021 |
| Isopentane | 0.052184 | 0.000027 | 0.049928 | 0.000021 |
| Neopentane | – | – | 0.049615 | 0.000031 |
| Hexane | 0.052567 | 0.000024 | 0.050708 | 0.000019 |
| Carbon dioxide | 1.622854 | 0.000302 | 4.001075 | 0.000283 |
| Nitrogen | 4.321699 | 0.000783 | 12.017829 | 0.000767 |
| Oxygen | 0.538012 | 0.000107 | – | – |
| Helium | – | – | 0.496897 | 0.000297 |
| Hydrogen | – | – | 3.009733 | 0.001323 |
| | | | | |
| *Hydrogen* | *0.000003* | *0.000002* | *–* | *–* |
| *Argon* | *0.000001* | *0.000001* | *–* | *–* |
| *Ethene* | *0.000003* | *0.000002* | *0.000001* | *< 0.000001* |
| *Neopentane* | *0.000019* | *0.000004* | *–* | *–* |
| *Carbon monoxide* | *0.000003* | *0.000003* | *0.000002* | *0.000002* |
| *Oxygen* | *–* | *–* | *0.000011* | *0.000006* |

**Table 3.** Results of the GC analysis and relative deviation between gravimetric composition and GC analysis for the two natural gas mixtures. The results are followed by the gravimetric composition of the employed validation mixtures (arranged in the same way as in Table 2).

| component | 11M (BAM-G 420) natural gas BAM no.: C49358-090825 | | | $H_2$-enriched natural gas CCQM K 118 BAM no.: 8099-160905 | | |
|---|---|---|---|---|---|---|
| | concentration | | relative deviation between gravimetric preparation and GC analysis | concentration | | relative deviation between gravimetric preparation and GC analysis |
| | $x_i$ / ($10^{-2}$ mol·mol$^{-1}$) | $U(x_i)$ ($k=2$) / ($10^{-2}$ mol·mol$^{-1}$) | % | $x_i$ / ($10^{-2}$ mol·mol$^{-1}$) | $U(x_i)$ ($k=2$) / ($10^{-2}$ mol·mol$^{-1}$) | % |
| Methane | 87.6348 | 0.0450 | –0.033 | 78.8143 | 0.0548 | –0.009 |
| Ethane | 4.2187 | 0.0086 | –0.155 | 0.7571 | 0.0046 | –0.036 |
| Propane | 1.0478 | 0.0022 | –0.113 | 0.2969 | 0.0009 | –0.045 |
| Butane | 0.2124 | 0.0002 | –0.131 | 0.2008 | 0.0002 | 0.172 |
| Isobutane | 0.2100 | 0.0002 | –0.135 | 0.1974 | 0.0002 | –0.292 |
| Pentane | 0.0518 | 0.0002 | –0.031 | 0.0501 | 0.0002 | –0.024 |
| Isopentane | 0.0522 | 0.0001 | –0.049 | 0.0499 | 0.0001 | 0.038 |
| Neopentane | – | – | – | 0.0495 | 0.0001 | –0.146 |
| Hexane | 0.0525 | 0.0003 | –0.106 | 0.0507 | 0.0001 | –0.048 |
| Carbon dioxide | 1.6242 | 0.0057 | 0.083 | 4.0005 | 0.0091 | –0.014 |
| Nitrogen | 4.3130 | 0.0017 | –0.200 | 12.0168 | 0.0037 | –0.008 |
| Oxygen | 0.5377 | 0.0005 | –0.062 | – | – | – |
| Helium | – | – | – | 0.4969 | 0.0007 | –0.005 |
| Hydrogen | – | – | – | 3.0105 | 0.0030 | 0.025 |
| | validation mixture BAM no.: 96054 925-090811 | | | validation mixture BAM no.: 8056-140922 | | |
| Methane | 89.021569 | 0.003176 | | 80.443530 | 0.003245 | |
| Ethane | 3.807373 | 0.000420 | | 0.700350 | 0.000141 | |
| Propane | 0.945272 | 0.001901 | | 0.274717 | 0.000080 | |
| Butane | 0.190906 | 0.000092 | | 0.184936 | 0.000091 | |

| | | | | |
|---|---|---|---|---|
| Isobutane | 0.190061 | 0.000077 | 0.186453 | 0.000035 |
| Pentane | 0.049116 | 0.000025 | 0.046127 | 0.000055 |
| Isopentane | 0.048774 | 0.000025 | 0.046163 | 0.000057 |
| Neopentane | – | – | 0.045306 | 0.000057 |
| Hexane | 0.047765 | 0.000022 | 0.046238 | 0.000083 |
| Carbon dioxide | 1.426743 | 0.000279 | 3.700223 | 0.000247 |
| Nitrogen | 3.799404 | 0.000718 | 11.103398 | 0.000655 |
| Oxygen | 0.472990 | 0.000097 | – | – |
| Helium | – | – | 0.471769 | 0.000280 |
| Hydrogen | – | – | 2.750777 | 0.001192 |
| | | | | |
| *Hydrogen* | *0.000003* | *0.000002* | – | – |
| *Argon* | *0.000001* | *0.000001* | – | – |
| *Ethene* | *0.000003* | *0.000002* | *0.000001* | *< 0.000001* |
| *Neopentane* | *0.000018* | *0.000004* | – | – |
| *Carbon monoxide* | *0.000003* | *0.000003* | *0.000002* | *0.000002* |
| *Oxygen* | – | – | *0.000011* | *0.000006* |

| | | |
|---|---|---|
| | validation mixture | |
| | BAM no.: 8045-140923 | |
| Methane | 77.208758 | 0.004338 |
| Ethane | 0.810092 | 0.000181 |
| Propane | 0.321367 | 0.000100 |
| Butane | 0.216555 | 0.000107 |
| Isobutane | 0.215044 | 0.000041 |
| Pentane | 0.054141 | 0.000064 |
| Isopentane | 0.054195 | 0.000067 |
| Neopentane | 0.052772 | 0.000067 |
| Hexane | 0.053205 | 0.000096 |
| Carbon dioxide | 4.304190 | 0.000318 |
| Nitrogen | 12.928316 | 0.000872 |
| Oxygen | – | – |
| Helium | 0.535826 | 0.000322 |
| Hydrogen | 3.245527 | 0.001441 |
| *Hydrogen* | *–* | *–* |
| *Argon* | *–* | *–* |
| *Ethene* | *0.000001* | *< 0.000001* |
| *Neopentane* | *–* | *–* |
| *Carbon monoxide* | *0.000002* | *0.000002* |
| *Oxygen* | *0.000011* | *0.000005* |

**Table 4.** Contributions to the expanded overall uncertainty in density $U_T (\rho_{exp})$ ($k = 2$) for the two studied natural gas mixtures in the temperature range from 260 to 350 K.

| Source of uncertainty | Units | Contribution ($k = 2$) | Estimation in density ($k = 2$) | |
|---|---|---|---|---|
| | | | kg·m$^{-3}$ | % |
| 11M (BAM-G 420) natural gas BAM no.: C49358-090825 | | | | |
| Temperature $T$ | K | 0.004 | < 0.008 | < 0.005 |
| Pressure $p$ | MPa | 0.005 | (0.002 – 0.069) | (0.015 – 0.191) |
| Density $\rho$ | kg·m$^{-3}$ | (0.023 – 0.049) | (0.023 – 0.049) | (0.022 – 0.378) |
| Composition $x_i$ | mol·mol$^{-1}$ | < 0.0004 | < 0.023 | (0.005 – 0.029) |
| Overall uncertainty $U_T (\rho_{exp})$ | | | (0.024 – 0.081) | (0.031 – 0.507) |
| H$_2$-enriched natural gas BAM no.: 8099-160905 | | | | |
| Temperature $T$ | K | 0.004 | < 0.007 | < 0.004 |
| Pressure $p$ | MPa | 0.005 | (0.002 – 0.049) | (0.006 – 0.189) |
| Density $\rho$ | kg·m$^{-3}$ | (0.024 – 0.049) | (0.024 – 0.049) | (0.022 – 0.372) |
| Composition $x_i$ | mol·mol$^{-1}$ | < 0.0004 | < 0.007 | < 0.003 |
| Overall uncertainty $U_T (\rho_{exp})$ | | | (0.024 – 0.071) | (0.029 – 0.502) |

**Table 5.** Experimental (p, ρ, T) measurements for the 11M (BAM-G 420) natural gas, relative and absolute expanded uncertainty in density (k = 2) $U(\rho_{exp})$, and relative deviations from the GERG-2008 and AGA8-DC92 EoS; where T is the temperature (ITS-90), p the pressure, $\rho_{exp}$ the experimental density, and $\rho_{GERG}$ and $\rho_{AGA}$ the densities calculated from the GERG-2008 and the AGA8-DC92 EoS.

| $T$/K [a] | $p$/MPa [a] | $\rho_{exp}$/kg·m$^{-3}$ | $U(\rho_{exp})$ / kg·m$^{-3}$ (k = 2) | $10^2\, U(\rho_{exp})/\rho_{exp}$ | $10^2\,(\rho_{exp} - \rho_{GERG})/\rho_{GERG}$ | $10^2\,(\rho_{exp} - \rho_{AGA})/\rho_{AGA}$ |
|---|---|---|---|---|---|---|
| 259.989 | 17.977 | 225.658 | 0.049 | 0.022 | −0.065 | −0.127 |
| 259.992 | 17.023 | 216.671 | 0.048 | 0.022 | −0.074 | −0.119 |
| 259.994 | 16.014 | 206.202 | 0.047 | 0.023 | −0.084 | −0.109 |
| 259.992 | 15.025 | 194.876 | 0.045 | 0.023 | −0.092 | −0.100 |
| 259.994 | 14.012 | 182.092 | 0.044 | 0.024 | −0.095 | −0.093 |
| 259.995 | 13.021 | 168.450 | 0.042 | 0.025 | −0.090 | −0.093 |
| 259.994 | 12.009 | 153.491 | 0.041 | 0.027 | −0.075 | −0.099 |
| 259.994 | 11.014 | 138.040 | 0.039 | 0.028 | −0.047 | −0.098 |
| 259.996 | 10.013 | 122.146 | 0.037 | 0.030 | −0.025 | −0.099 |
| 259.990 | 9.008 | 106.300 | 0.035 | 0.033 | −0.007 | −0.096 |
| 259.993 | 7.988 | 90.764 | 0.034 | 0.037 | 0.010 | - |
| 259.991 | 7.000 | 76.503 | 0.032 | 0.042 | 0.020 | −0.077 |
| 259.992 | 5.999 | 62.980 | 0.030 | 0.048 | 0.019 | −0.075 |
| 259.992 | 4.998 | 50.417 | 0.029 | 0.057 | 0.015 | −0.072 |
| 259.993 | 4.002 | 38.839 | 0.028 | 0.071 | 0.008 | −0.068 |
| 259.994 | 2.997 | 28.021 | 0.026 | 0.094 | 0.005 | −0.059 |
| 259.992 | 1.992 | 17.972 | 0.025 | 0.140 | −0.001 | −0.048 |
| 259.991 | 0.998 | 8.701 | 0.024 | 0.278 | −0.013 | −0.041 |
| 274.977 | 19.622 | 213.256 | 0.048 | 0.022 | −0.057 | −0.117 |
| 274.977 | 19.015 | 208.165 | 0.047 | 0.023 | −0.061 | −0.114 |
| 274.975 | 17.999 | 199.103 | 0.046 | 0.023 | −0.067 | −0.108 |
| 274.976 | 17.010 | 189.597 | 0.045 | 0.024 | −0.071 | −0.102 |
| 274.976 | 16.002 | 179.179 | 0.044 | 0.024 | −0.072 | −0.099 |
| 274.973 | 14.994 | 168.038 | 0.042 | 0.025 | −0.071 | −0.099 |
| 274.971 | 14.005 | 156.441 | 0.041 | 0.026 | −0.065 | −0.100 |
| 274.972 | 13.007 | 144.175 | 0.040 | 0.027 | −0.056 | −0.102 |
| 274.972 | 12.009 | 131.463 | 0.038 | 0.029 | −0.044 | −0.100 |
| 274.971 | 11.007 | 118.489 | 0.037 | 0.031 | −0.027 | −0.092 |
| 274.972 | 10.004 | 105.462 | 0.035 | 0.033 | −0.018 | −0.090 |
| 274.971 | 9.001 | 92.621 | 0.034 | 0.036 | −0.010 | −0.087 |
| 274.971 | 8.000 | 80.158 | 0.032 | 0.040 | −0.005 | −0.085 |
| 274.971 | 7.001 | 68.196 | 0.031 | 0.045 | 0.001 | −0.079 |
| 274.972 | 5.998 | 56.751 | 0.030 | 0.052 | 0.003 | −0.075 |
| 274.972 | 4.997 | 45.910 | 0.028 | 0.062 | 0.002 | −0.070 |
| 274.972 | 4.001 | 35.703 | 0.027 | 0.076 | −0.001 | −0.064 |
| 274.971 | 3.000 | 26.006 | 0.026 | 0.101 | −0.003 | −0.054 |
| 274.968 | 2.000 | 16.860 | 0.025 | 0.149 | < 0.001 | −0.038 |
| 274.969 | 0.998 | 8.183 | 0.024 | 0.295 | 0.001 | −0.021 |
| 299.934 | 16.303 | 150.195 | 0.040 | 0.027 | −0.027 | −0.072 |
| 299.934 | 16.008 | 147.459 | 0.040 | 0.027 | −0.026 | −0.072 |
| 299.933 | 15.003 | 137.934 | 0.039 | 0.028 | −0.022 | −0.073 |
| 299.932 | 13.999 | 128.156 | 0.038 | 0.030 | −0.017 | −0.073 |
| 299.933 | 13.000 | 118.212 | 0.037 | 0.031 | −0.012 | −0.072 |
| 299.934 | 12.000 | 108.142 | 0.036 | 0.033 | −0.008 | −0.072 |
| 299.932 | 11.001 | 98.044 | 0.034 | 0.035 | 0.001 | −0.065 |
| 299.931 | 10.001 | 87.967 | 0.033 | 0.038 | 0.002 | −0.066 |
| 299.930 | 9.001 | 78.002 | 0.032 | 0.041 | 0.003 | −0.065 |
| 299.931 | 7.999 | 68.196 | 0.031 | 0.045 | 0.004 | −0.064 |
| 299.930 | 6.998 | 58.628 | 0.030 | 0.051 | 0.004 | −0.060 |
| 299.926 | 5.996 | 49.322 | 0.029 | 0.058 | 0.004 | −0.056 |
| 299.927 | 4.998 | 40.335 | 0.028 | 0.069 | 0.004 | −0.049 |
| 299.926 | 3.998 | 31.645 | 0.027 | 0.085 | 0.001 | −0.044 |
| 299.928 | 2.997 | 23.266 | 0.026 | 0.111 | 0.003 | −0.033 |
| 299.926 | 1.997 | 15.204 | 0.025 | 0.164 | 0.002 | −0.025 |
| 299.923 | 0.997 | 7.446 | 0.024 | 0.323 | −0.005 | −0.021 |
| 324.932 | 18.569 | 146.171 | 0.040 | 0.027 | −0.047 | −0.099 |
| 324.934 | 18.000 | 141.876 | 0.039 | 0.028 | −0.046 | −0.100 |
| 324.936 | 16.998 | 134.140 | 0.038 | 0.029 | −0.044 | −0.102 |
| 324.936 | 15.997 | 126.218 | 0.038 | 0.030 | −0.042 | −0.104 |
| 324.936 | 14.997 | 118.150 | 0.037 | 0.031 | −0.038 | −0.103 |
| 324.937 | 13.998 | 109.947 | 0.036 | 0.033 | −0.036 | −0.102 |
| 324.937 | 12.999 | 101.654 | 0.035 | 0.034 | −0.034 | −0.101 |
| 324.937 | 11.998 | 93.287 | 0.034 | 0.036 | −0.031 | −0.099 |
| 324.936 | 10.997 | 84.908 | 0.033 | 0.039 | −0.024 | −0.090 |
| 324.936 | 10.001 | 76.578 | 0.032 | 0.042 | −0.027 | −0.091 |
| 324.935 | 8.998 | 68.248 | 0.031 | 0.045 | −0.025 | −0.087 |
| 324.933 | 7.997 | 60.023 | 0.030 | 0.050 | −0.023 | −0.082 |
| 324.933 | 6.997 | 51.928 | 0.029 | 0.056 | −0.021 | −0.075 |
| 324.934 | 5.997 | 43.970 | 0.028 | 0.064 | −0.022 | −0.070 |
| 324.934 | 4.997 | 36.174 | 0.027 | 0.076 | −0.023 | −0.065 |
| 324.934 | 3.997 | 28.555 | 0.026 | 0.093 | −0.022 | −0.057 |
| 324.934 | 2.997 | 21.122 | 0.026 | 0.121 | −0.023 | −0.051 |
| 324.930 | 1.990 | 13.831 | 0.025 | 0.179 | −0.023 | −0.044 |
| 324.929 | 0.998 | 6.837 | 0.024 | 0.351 | −0.032 | −0.044 |
| 349.919 | 19.701 | 136.343 | 0.039 | 0.028 | –0.031 | –0.095 |
| 349.918 | 18.999 | 131.749 | 0.038 | 0.029 | –0.030 | –0.096 |
| 349.918 | 17.997 | 125.065 | 0.037 | 0.030 | –0.029 | –0.098 |
| 349.919 | 16.995 | 118.246 | 0.037 | 0.031 | –0.027 | –0.098 |
| 349.919 | 15.997 | 111.336 | 0.036 | 0.032 | –0.026 | –0.098 |
| 349.921 | 14.996 | 104.305 | 0.035 | 0.034 | –0.024 | –0.096 |
| 349.920 | 13.996 | 97.202 | 0.034 | 0.035 | –0.023 | –0.094 |
| 349.919 | 12.995 | 90.022 | 0.033 | 0.037 | –0.022 | –0.092 |
| 349.919 | 11.997 | 82.823 | 0.033 | 0.039 | –0.021 | –0.089 |
| 349.918 | 10.999 | 75.612 | 0.032 | 0.042 | –0.018 | –0.082 |
| 349.919 | 9.997 | 68.368 | 0.031 | 0.045 | –0.019 | –0.079 |
| 349.919 | 8.996 | 61.155 | 0.030 | 0.049 | –0.020 | –0.076 |
| 349.919 | 7.996 | 53.989 | 0.029 | 0.054 | –0.021 | –0.072 |
| 349.918 | 6.997 | 46.892 | 0.029 | 0.061 | –0.019 | –0.065 |
| 349.918 | 5.996 | 39.864 | 0.028 | 0.070 | –0.021 | –0.061 |
| 349.918 | 4.997 | 32.935 | 0.027 | 0.082 | –0.022 | –0.057 |
| 349.918 | 3.996 | 26.095 | 0.026 | 0.100 | –0.025 | –0.055 |
| 349.918 | 2.998 | 19.395 | 0.025 | 0.131 | –0.026 | –0.050 |
| 349.917 | 1.997 | 12.791 | 0.025 | 0.193 | –0.020 | –0.037 |
| 349.917 | 0.997 | 6.323 | 0.024 | 0.378 | –0.031 | –0.042 |

[a] Expanded uncertainties ($k = 2$) in temperature and pressure are $U(T) = 0.004$ K and $U(p) = 0.005$ MPa, respectively.

**Table 6.** Experimental (p, ρ, T) measurements for the H$_2$-enriched natural gas mixture, relative and absolute expanded uncertainty in density (k = 2) U(ρ$_{exp}$), and relative deviations from the GERG-2008 and AGA8-DC92 EoS; where T is the temperature (ITS-90), p the pressure, ρ$_{exp}$ the experimental density, and ρ$_{GERG}$ and ρ$_{AGA}$ the densities calculated from the GERG-2008 and the AGA8-DC92 EoS.

| T/K [a] | p/MPa [a] | ρ$_{exp}$/kg·m$^{-3}$ | U(ρ$_{exp}$) / kg·m$^{-3}$ (k = 2) | 10$^2$ U(ρ$_{exp}$) /ρ$_{exp}$ | 10$^2$ (ρ$_{exp}$−ρ$_{GERG}$)/ρ$_{GERG}$ | 10$^2$ (ρ$_{exp}$−ρ$_{AGA}$)/ρ$_{AGA}$ |
|---|---|---|---|---|---|---|
| 260.041 | 19.843 | 224.730 | 0.049 | 0.022 | −0.254 | −0.193 |
| 260.038 | 19.011 | 217.428 | 0.048 | 0.022 | −0.268 | −0.178 |
| 260.040 | 18.010 | 208.053 | 0.047 | 0.023 | −0.280 | −0.156 |
| 260.041 | 17.008 | 197.993 | 0.046 | 0.023 | −0.288 | −0.133 |
| 260.040 | 16.010 | 187.277 | 0.045 | 0.024 | −0.291 | −0.114 |
| 260.040 | 15.012 | 175.878 | 0.043 | 0.025 | −0.285 | −0.097 |
| 260.040 | 14.011 | 163.801 | 0.042 | 0.026 | −0.269 | −0.083 |
| 260.039 | 13.012 | 151.191 | 0.040 | 0.027 | −0.245 | −0.071 |
| 260.038 | 12.011 | 138.130 | 0.039 | 0.028 | −0.214 | −0.063 |
| 260.039 | 11.009 | 124.803 | 0.037 | 0.030 | −0.171 | −0.051 |
| 260.041 | 10.008 | 111.413 | 0.036 | 0.032 | −0.135 | −0.050 |
| 260.040 | 9.005 | 98.132 | 0.034 | 0.035 | −0.104 | - |
| 260.041 | 8.010 | 85.247 | 0.033 | 0.039 | −0.075 | −0.059 |
| 260.040 | 7.007 | 72.690 | 0.031 | 0.043 | −0.050 | −0.060 |
| 260.040 | 6.001 | 60.600 | 0.030 | 0.050 | −0.035 | −0.064 |
| 260.038 | 4.999 | 49.124 | 0.029 | 0.059 | −0.025 | −0.064 |
| 260.039 | 4.001 | 38.246 | 0.028 | 0.072 | −0.018 | −0.060 |
| 260.040 | 2.993 | 27.840 | 0.026 | 0.095 | < 0.001 | −0.039 |
| 260.041 | 1.998 | 18.098 | 0.025 | 0.140 | −0.013 | −0.045 |
| 260.041 | 0.999 | 8.815 | 0.024 | 0.275 | 0.001 | −0.019 |
| | | | | | | |
| 275.020 | 19.234 | 196.179 | 0.046 | 0.023 | −0.225 | −0.126 |
| 275.019 | 18.008 | 185.120 | 0.044 | 0.024 | −0.228 | −0.111 |
| 275.018 | 17.013 | 175.624 | 0.043 | 0.025 | −0.226 | −0.101 |
| 275.018 | 16.006 | 165.559 | 0.042 | 0.025 | −0.220 | −0.092 |
| 275.016 | 15.004 | 155.107 | 0.041 | 0.026 | −0.209 | −0.084 |
| 275.017 | 14.008 | 144.361 | 0.040 | 0.027 | −0.193 | −0.076 |
| 275.016 | 13.007 | 133.256 | 0.038 | 0.029 | −0.174 | −0.070 |
| 275.015 | 12.006 | 121.941 | 0.037 | 0.030 | −0.155 | −0.067 |
| 275.015 | 11.004 | 110.531 | 0.036 | 0.032 | −0.128 | −0.060 |
| 275.013 | 10.006 | 99.141 | 0.035 | 0.035 | −0.111 | −0.063 |
| 275.014 | 9.003 | 87.821 | 0.033 | 0.038 | −0.092 | −0.065 |
| 275.013 | 8.002 | 76.705 | 0.032 | 0.042 | −0.076 | −0.069 |
| 275.013 | 7.001 | 65.854 | 0.031 | 0.047 | −0.060 | −0.068 |
| 275.013 | 6.000 | 55.328 | 0.030 | 0.053 | −0.049 | −0.067 |
| 275.013 | 5.003 | 45.205 | 0.028 | 0.063 | −0.038 | −0.062 |
| 275.012 | 4.006 | 35.445 | 0.027 | 0.077 | −0.027 | −0.054 |
| 275.011 | 3.001 | 26.000 | 0.026 | 0.101 | −0.012 | −0.037 |
| 275.011 | 2.000 | 16.957 | 0.025 | 0.148 | −0.003 | −0.025 |
| 275.011 | 0.999 | 8.296 | 0.024 | 0.291 | 0.009 | −0.005 |
| | | | | | | |
| 299.961 | 19.851 | 171.675 | 0.043 | 0.025 | −0.162 | −0.089 |
| 299.961 | 19.009 | 165.029 | 0.042 | 0.025 | −0.161 | −0.086 |
| 299.960 | 18.003 | 156.843 | 0.041 | 0.026 | −0.157 | −0.084 |
| 299.960 | 17.002 | 148.434 | 0.040 | 0.027 | −0.152 | −0.081 |
| 299.960 | 16.003 | 139.800 | 0.039 | 0.028 | −0.145 | −0.078 |
| 299.960 | 15.002 | 130.945 | 0.038 | 0.029 | −0.137 | −0.075 |
| 299.960 | 14.002 | 121.924 | 0.037 | 0.030 | −0.127 | −0.072 |
| 299.960 | 13.003 | 112.776 | 0.036 | 0.032 | −0.116 | −0.069 |
| 299.958 | 12.003 | 103.536 | 0.035 | 0.034 | −0.108 | −0.069 |
| 299.958 | 11.000 | 94.230 | 0.034 | 0.036 | −0.091 | −0.061 |
| 299.958 | 10.002 | 84.966 | 0.033 | 0.039 | −0.082 | −0.062 |
| 299.958 | 9.001 | 75.735 | 0.032 | 0.042 | −0.071 | −0.059 |
| 299.957 | 8.000 | 66.591 | 0.031 | 0.046 | −0.061 | −0.057 |
| 299.956 | 7.002 | 57.603 | 0.030 | 0.052 | −0.049 | −0.052 |
| 299.956 | 5.999 | 48.731 | 0.029 | 0.059 | −0.042 | −0.049 |
| 299.957 | 4.999 | 40.073 | 0.028 | 0.069 | −0.031 | −0.042 |
| 299.959 | 3.998 | 31.613 | 0.027 | 0.085 | −0.021 | −0.033 |
| 299.958 | 2.991 | 23.309 | 0.026 | 0.111 | −0.008 | −0.020 |
| 299.957 | 1.999 | 15.356 | 0.025 | 0.162 | −0.004 | −0.016 |
| 299.958 | 0.999 | 7.557 | 0.024 | 0.318 | 0.001 | −0.007 |
| | | | | | | |
| 324.963 | 19.928 | 150.681 | 0.040 | 0.027 | −0.107 | −0.064 |
| 324.962 | 19.001 | 144.143 | 0.040 | 0.027 | −0.105 | −0.065 |
| 324.963 | 17.999 | 136.914 | 0.039 | 0.028 | −0.102 | −0.064 |
| 324.963 | 17.000 | 129.541 | 0.038 | 0.029 | −0.098 | −0.063 |
| 324.964 | 16.000 | 122.027 | 0.037 | 0.030 | −0.094 | −0.061 |
| 324.963 | 14.998 | 114.372 | 0.036 | 0.032 | −0.088 | −0.059 |
| 324.963 | 14.001 | 106.647 | 0.035 | 0.033 | −0.082 | −0.056 |
| 324.963 | 13.002 | 98.826 | 0.034 | 0.035 | −0.076 | −0.054 |
| 324.962 | 12.001 | 90.934 | 0.034 | 0.037 | −0.069 | −0.051 |
| 324.963 | 11.000 | 83.015 | 0.033 | 0.039 | −0.057 | −0.042 |
| 324.963 | 9.999 | 75.089 | 0.032 | 0.042 | −0.053 | −0.042 |
| 324.962 | 8.999 | 67.186 | 0.031 | 0.046 | −0.043 | −0.036 |
| 324.962 | 7.999 | 59.321 | 0.030 | 0.051 | −0.037 | −0.033 |
| 324.962 | 6.999 | 51.523 | 0.029 | 0.056 | −0.028 | −0.027 |
| 324.962 | 5.998 | 43.796 | 0.028 | 0.064 | −0.023 | −0.025 |
| 324.961 | 4.999 | 36.180 | 0.027 | 0.076 | −0.018 | −0.022 |
| 324.962 | 3.998 | 28.673 | 0.026 | 0.092 | –0.012 | –0.018 |
| 324.961 | 2.985 | 21.199 | 0.026 | 0.121 | –0.002 | –0.009 |
| 324.961 | 1.999 | 14.050 | 0.025 | 0.177 | –0.001 | –0.008 |
| 324.962 | 0.998 | 6.945 | 0.024 | 0.345 | 0.005 | –0.001 |
| 349.946 | 19.941 | 134.571 | 0.039 | 0.029 | –0.110 | –0.091 |
| 349.946 | 19.000 | 128.613 | 0.038 | 0.029 | –0.108 | –0.090 |
| 349.945 | 17.997 | 122.149 | 0.037 | 0.030 | –0.106 | –0.089 |
| 349.946 | 16.995 | 115.580 | 0.036 | 0.031 | –0.102 | –0.087 |
| 349.944 | 15.998 | 108.944 | 0.036 | 0.033 | –0.099 | –0.085 |
| 349.946 | 14.996 | 102.190 | 0.035 | 0.034 | –0.094 | –0.081 |
| 349.946 | 13.998 | 95.389 | 0.034 | 0.036 | –0.089 | –0.077 |
| 349.945 | 12.998 | 88.509 | 0.033 | 0.038 | –0.085 | –0.074 |
| 349.944 | 11.991 | 81.546 | 0.032 | 0.040 | –0.073 | –0.063 |
| 349.945 | 10.996 | 74.624 | 0.032 | 0.042 | –0.064 | –0.056 |
| 349.945 | 9.997 | 67.658 | 0.031 | 0.046 | –0.062 | –0.055 |
| 349.946 | 8.998 | 60.691 | 0.030 | 0.050 | –0.057 | –0.052 |
| 349.947 | 7.998 | 53.731 | 0.029 | 0.055 | –0.052 | –0.049 |
| 349.947 | 6.997 | 46.792 | 0.029 | 0.061 | –0.045 | –0.044 |
| 349.947 | 5.998 | 39.899 | 0.028 | 0.070 | –0.041 | –0.042 |
| 349.946 | 4.998 | 33.064 | 0.027 | 0.082 | –0.037 | –0.040 |
| 349.946 | 3.998 | 26.284 | 0.026 | 0.100 | –0.032 | –0.037 |
| 349.946 | 2.985 | 19.492 | 0.025 | 0.130 | –0.026 | –0.031 |
| 349.946 | 1.999 | 12.964 | 0.025 | 0.190 | –0.025 | –0.031 |
| 349.946 | 0.999 | 6.431 | 0.024 | 0.372 | –0.007 | –0.012 |

[a] Expanded uncertainties ($k = 2$) in temperature and pressure are $U(T) = 0.004$ K and $U(p) = 0.005$ MPa, respectively.

**Table 7**. Statistical parameters of the ($p$, $\rho$, $T$) data set with respect to the GERG-2008 and AGA8-DC92 EoS for the two studied natural gas mixtures including literature data for comparable mixtures.

| reference | identifier | covered ranges | | $N$ | experimental vs. GERG-2008 | | | | experimental vs. AGA8-DC92 | | | |
|---|---|---|---|---|---|---|---|---|---|---|---|---|
| | | $T$ / K | $p$ / MPa | | AAD | Bias | RMS | MaxD / % | AAD | Bias | RMS | MaxD / % |
| this work (2018) | 11M natural gas | 260 – 350 | 1 – 20 | 94 | 0.027 | –0.025 | 0.036 | 0.095 | 0.078 | –0.078 | 0.081 | 0.127 |
| this work (2018) | $H_2$-enriched natural gas | 260 – 350 | 1 – 20 | 99 | 0.095 | –0.095 | 0.123 | 0.291 | 0.062 | –0.062 | 0.070 | 0.193 |
| Richter et al. (2014) [30] | NG1 | 273 – 293 | 1.0 – 8.0 | 37 | 0.021 | 0.018 | 0.023 | 0.0375 | 0.010 | 0.001 | 0.013 | 0.0338 |
| Richter et al. (2014) [30] | NG2 | 273 – 293 | 1.0 – 8.0 | 36 | 0.032 | 0.032 | 0.036 | 0.0664 | 0.013 | 0.004 | 0.018 | 0.0509 |
| Richter et al. (2014) [30] | NG3 | 283 | 1.0 – 8.0 | 13 | 0.014 | 0.009 | 0.015 | 0.0256 | 0.027 | –0.026 | 0.031 | 0.0447 |
| Atilhan et al. (2015) [31] | M88C1 | 270 – 340 | 3.5 – 34.5 | 32 | 0.221 | –0.045 | 0.261 | 0.639 | 0.365 | 0.299 | 0.384 | 0.610 |
| Atilhan et al. (2015) [31] | M94C1 | 270 – 340 | 3.5 – 34.5 | 61 | 0.186 | –0.144 | 0.215 | 0.516 | 0.094 | 0.039 | 0.112 | 0.361 |
| Ahmadi et al. (2017) [32] | natural gas | 323 – 415 | 1.3 – 58.4 | 110 | 0.135 | –0.002 | 0.303 | 2.18 | 0.1[a] | | | |

[a] Reference 32 does not provide tabled individual data of the deviation resulting from the AGA8-DC92 EoS.